\documentclass[a4paper, 10pt]{article}

\usepackage{authblk}        
\usepackage{comment}        

\usepackage{amssymb}        
\usepackage{mathtools}  

\usepackage{graphicx} 
\usepackage{float}   
\usepackage{caption}  

\usepackage{amsthm}         
\theoremstyle{plain}
\newtheorem{thm}{Theorem}[section] 

\theoremstyle{definition}
\newtheorem{defn}[thm]{Definition} 

\usepackage{graphicx}  

\usepackage{tikz-cd}
\usetikzlibrary{backgrounds}      

\tikzset{
  boxed/.style={                  
    show background rectangle,    
    background rectangle/.append style={ 
      draw=black, thick, rounded corners}}}
      
\begin{document}

\begin{titlepage}
	
\title{A Category of Genes}
	
\author[1,2]{\small Yanying Wu\thanks{yanying.wu@cncb.ox.ac.uk}}
\affil[1]{\small Centre for Neural Circuits and Behaviour, University of Oxford, UK}
\affil[2]{\small Department of Physiology, Anatomy and Genetics, University of Oxford, UK}
\date{12 Nov, 2023}
\clearpage\maketitle
\thispagestyle{empty}
\vspace{5mm}	
	
\begin{abstract}

Understanding how genes interact and relate to each other is a fundamental question in biology. However, current practices for describing these relationships, such as drawing diagrams or graphs in a somewhat arbitrary manner, limit our ability to integrate various aspects of the gene functions and view the genome holistically. To overcome these limitations, we need a more appropriate way to describe the intricate relationships between genes.

Interestingly, category theory, an abstract field of mathematics seemingly unrelated to biology, has emerged as a powerful language for describing relations in general. We propose that category theory could provide a framework for unifying our knowledge of genes and their relationships.

As a starting point, we construct a category of genes, with its morphisms abstracting various aspects of the relationships betweens genes. These relationships include, but not limited to, the order of genes on the chromosomes, the physical or genetic interactions, the signalling pathways, the gene ontology causal activity models (GO-CAM) and gene groups. Previously, they were encoded by miscellaneous networks or graphs, while our work unifies them in a consistent manner as a category. By doing so, we hope to view the relationships between genes systematically. In the long run, this paves a promising way for us to understand the fundamental principles that govern gene regulation and function.

\end{abstract}
\end{titlepage}

\section{Introduction} \label{introduction}
In recent years, the amount of available genomic data has increased exponentially, making it possible to study the complex interactions between genes in great details. One of the main challenges in this field is developing new mathematical and computational tools that can accurately capture and analyze the relationships between genes. In this regard, category theory has emerged as a promising framework for modelling complex systems.

Category theory is a branch of mathematics that provides a powerful tool for studying relationships between objects. The power of category theory lies in its ability to identify deep similarities among the structures of different entities, and therefore to reveal their relationships that are undetectable at the superficial layer \cite{sica2006category, Southwell2021}. This power has extended the application of category theory beyond pure mathematics, and even to form its own field. Applied category theory is an developing field that aims to apply the concepts, methods, and tools of category theory to various areas of science, engineering, and technology. Thus far, there has been a growing interest in using category theory to solve problems in fields such as computer science \cite{Milewski2019}, quantum mechanism \cite{Coecke2017}, dynamic systems \cite{fong2018seven} and so on. In particular, applied category theory has been used to study genetics \cite{Tuyeras2018, Tuyeras2018a}. However, the potential of applied category theory in genomics research has not been fully appreciated.

In this study, we aim to explore the use of category theory to describe the relationships between genes, which is one of the greatest mysteries in genomics. We will first review several potential categorical representations of genes, such as a preorder for gene orders \cite{Wu2020}, the open Petri nets for gene regulatory networks \cite{Wu2019}, an olog for gene ontology \cite{Wu2019b} and an operad for gene trees \cite{Baez2015}.

Then, we will introduce a more abstract category for genes: the $\textbf{Dist}$-category enriched in a symmetric monoidal preorder $(\mathbb{N}, \ge, 0, +)$ \cite{fong2018seven}. Here, we define the objects to be genes and the morphisms to be distances. A distance between two genes is a natural number. This category is a template category that can be instantiated into various contexts. For example, in the preorder scenario mentioned above, we define the distance between two genes $g_1$ and $g_2$ as $1$, if $g_1$ is the closest precedent of $g_2$ on the linear form of a specific genome. Similarly, distances could represent different relations in other situations, which we will elaborate in turn. 


\section{Candidate categories for the genes} \label{catgene_review}
\subsection{What is a category?}
A category is basically a collection of objects and their relationships. In terms of mathematical definition, we have \cite{awodey2010category, riehl2017category, spivak2014category} :
\begin{defn}
A $\textit{category}$ $\mathbb{C}$ consists of the following data:

\begin{itemize}
\item a collection of objects $\mathrm{Ob(\mathbb{C})}: x, y, z, ...$
\item for every pair $x, y \in \mathrm{Ob}(\mathbb{C})$, a set of morphisms $\mathrm{Hom}_\mathbb{C}(x, y)$ called hom-set from $x$ to $y$, for each morphism $f: x \rightarrow y \in \mathrm{Hom}_\mathbb{C}(x, y)$, $x$ is called the \textit{domain} and $y$ the \textit{codomain} of \textit{f}.

\item given morphisms $f: x \rightarrow y$ and $g: y \rightarrow z$, that is, the domain of \textit{g} is the same as the codomain of \textit{f}, there is a morphism $g \circ f: x \rightarrow z$ called the \textit{composite} of \textit{f} and \textit{g}.
\item for each object \textit{x}, there is a given morphism $id_x: x \rightarrow x$ called the \textit{identity morphism} of \textit{x}.
\end{itemize}

These data are required to satisfy the following laws:

\begin{itemize}
\item Associativity: $h \circ (g \circ f) = (h \circ g) \circ f$ for all $f: x \rightarrow y, \hspace*{1em} g: y \rightarrow z, \hspace*{1em} h: z \rightarrow w$.
\item Unit: $f \circ 1_x = f = 1_y \circ f$ for all $f: x \rightarrow y$. \hfill\(\Diamond\)
\end{itemize} 
\end{defn}

\noindent
Based on the definition of category, it is most straightforward to treat genes as objects, when we construct a category of genes. However, the tricky part is how we define the morphisms between genes. According to what we know from biological research, there are multiple ways to describe the relationships between genes. For example, each gene has its own coordinate on the chromosome, and we could order the genes according to their positions. Therefore, a linear ordered relationship between genes can be established. Moreover, when we consider the functions of genes, there exist important inter-gene interactions that form gene regulatory networks or signalling pathways. These relationships could be modelled in different categories of genes. 

In the following subsections, we review various candidate categories of genes, beginning with a preorder.

\subsection{Using a preorder to represent gene orders} \label{preorder}
\label{preorder1}
\subsubsection{Definitions of preorder, partial order and linear order}

Preorder is a basic and simple category. By definition \cite{spivak2014category}, we have:

\begin{defn}
A $\textit{preorder}$ refers to a set $S$ and a binary relation $\leq$ on $S$, paired and denoted as $(S, \leq)$, and it has the following properties:

\begin{itemize}
\item Reflexivity: $x \leq x$ for any $x \in S$.
\item Transitivity: if $x \leq y$ and $y \leq z$, then $x \leq z$, for any $x, y, z \in S$. \hfill\(\Diamond\)
\end{itemize} 
\end{defn}

\noindent
It is not difficult to see that $(S, \leq)$ is a category, and we denote it as $\mathbb{S}$. Comparing to the definition of a category,  the set $S$ is exactly the set of objects $\mathrm{Ob}(\mathbb{S})$, and the binary relations between objects are the morphisms. Further, transitivity of $(S, \leq)$ is equivalent to composition, and it is associative. Finally, reflexivity manifests the unit law.

Based on preorder, we define partial order \cite{fong2018seven}:
\begin{defn}
A $\textit{partial order}$ is a preorder that satisfies the following condition:
\begin{itemize}
\item Antisymmetry: If $x \leq y$ and $y \leq x$, then $x = y$. \hfill\(\Diamond\)
\end{itemize}
\end{defn}

\noindent
Partial order is a bit more restricted than preorder. Further, we have linear order \cite{fong2018seven}:
\begin{defn}
A $\textit{linear order}$ is a partial order which satisfies the following condition:
\begin{itemize}
\item Comparability: for all $x, y \in S$, either $x \leq y$ or $y \leq x$. \hfill\(\Diamond\)
\end{itemize} 
\end{defn}

\noindent
Intuitively, if we consider the graphic representation of a category, in which objects are vertices and morphisms are arrows between them, then we can think of a partial order as a preorder without loops, and a linear order as a partial order that has a path connecting any two vertices.

Note that linear order is also called total order \cite{fong2018seven}.

\subsubsection{The order of genes}
The genome of each and every species on the earth can be viewed as a set of linear DNA sequences taking the form of chromosomes. The chromosomes are conventionally named in sequential order, with the X and Y chromosomes placed at the end.

Besides, a gene occupies a fragment of DNA sequence, and it has definite start and stop coordinates on a certain chromosome. 

The linear and sequential nature of gene positions enables the ordering of any two different genes. 
If two genes are located on the same chromosome, the positions of these genes can be compared, or ordered according to their coordinates. 
While if they are located on different chromosomes, the order can be determined by that of their respective chromosomes.

\subsubsection{A category of genes as a preorder}

The first category of genes, a preorder, was presented in \cite{Wu2020}. Here, we will restate and exemplify its definition with more details.

We can construct a dedicated category $\mathbb{G}$ for each specific genome. The objects of this category are all the genes encoded in the genome, denoted as $\mathrm{Ob}(\mathbb{G})$. In order to define morphisms of $\mathbb{G}$, we need to introduce a few more concepts. 

Suppose our genome of interest has a set of chromosomes denoted as $\mathrm{Chr}(\mathbb{G})$, the elements of which are ordered according to conventional chromosome labels. We define function $ chr:  \mathrm{Ob}(\mathbb{G}) \rightarrow \mathrm{Chr}(\mathbb{G})$ to map a gene to its chromosome label, and functions $start, stop: \mathrm{Ob}(\mathbb{G}) \rightarrow \mathbb{N}$ to map a gene to its start and end coordinates, respectively.

We are now ready for the definition of morphism of $\mathbb{G}$, the preorder relationship between genes, denoted as $\leq$, and the definition of category $\mathbb{G}$ itself.
\begin{defn}
A $\textit{category of genes}$ $\mathbb{G}$ for a genome, in the form of a preorder, consists of the following data:
\begin{itemize}
\item a collection of genes $\mathrm{Ob}(\mathbb{G})$ that are encoded by the genome
\item for any two genes $x, y \in \mathrm{Ob}(\mathbb{G}) (x \neq y) $, there exists a preorder relation $x \leq y$ if one of the following conditions is met:
\begin{itemize}
\renewcommand{\labelenumii}{\arabic{enumi}.\arabic{enumii}}
\item $chr(x) = chr(y)$ and $start(x) < start(y)$
\item $chr(x) = chr(y)$ and $start(x) = start(y)$ and $stop(x) < stop(y)$
\item $chr(x) \neq chr(y)$ and $chr(x)$ is ordered in front of $chr(y)$
\end{itemize}

in addition, it satisfies the following properties:
\begin{itemize}
\item Reflexivity: $x \leq x$ for any $x \in \mathbb{G}$.
\item Transitivity: if $x \leq y$ and $y \leq z$, then $x \leq z$, for any $x, y, z \in \mathrm{Ob}(\mathbb{G})$. \hfill\(\Diamond\)
\end{itemize}
\end{itemize}
\end{defn}  

\noindent
Moreover, for any pairs of genes $g1, g2 \in \mathrm{Ob}(\mathbb{G})$, we have either $g1 \leq g2$ or $g2 \leq g1$, and if $g1 \leq g2$ and $g2 \leq g1$, then $g1=g2$. Therefore, the preorder of genes $\mathbb{G}$ is both a partial order and a linear (total) order.

\subsection{Using an open Petri net to represent a gene regulatory network} 
\label{petri}

The preorder of genes considers only the relative positions of genes but not their functions. Currently, a connection between the function of a gene and its position is unavailable. More common ways to describe the functional relationships between genes include gene regulatory network, signalling pathway, and gene ontology. We will cover the concept of a gene regulatory network and its categorification as an open Petri net in this subsection.

A gene regulatory network (GRN) is a set of genes, or parts of genes, that interact with each other to control a specific cell function. GRNs are important in development, differentiation and responding to environmental cues (https://www.nature.com/subjects/gene-regulatory-networks). A GRN is typically represented by a graph of network, where the nodes denote genes and the edges between two nodes denote their regulatory relations. The genes in a GRN normally consist of two main types: those encode transcription factors (TFs) and those encode the targets of the TFs. As for the regulatory relations, they include binding, inhibition, promoting, etc. 

Although a representation of GRNs as a network captures our knowledge of the biological reality to a large extend, it is informal and therefore are not suitable for an in-depth study of GRNs. To address this issue, several computational methods for modelling GRNs have been developed \cite{Karlebach2008}. We focus on the Petri net in particular due to its correlation with open Petri net, which is a categorical treatment of networks.

Petri net is a bipartite directed graph that has been proved to be useful in modelling distributed and concurrent systems \cite{petri1962kommunikation, Reisig1985, Murata1989}. When introduced to model GRNs, Petri net showed its advantage in analysis of the dynamics of the regulatory networks \cite{Steggles2007, Karlebach2008, Bordon2012}. On top of that, an open Petri net equips Petri net with two interface sets to allow tokens to flow in and out of the Petri net, thus makes it ``open". Following is a formal definition of open Petri net \cite{Baez2020}:
\begin{defn}
An $\textit{open Petri net}$ is a diagram in Petri of the form 

\begin{figure}[H]
	\centering
	\includegraphics[scale=0.3]{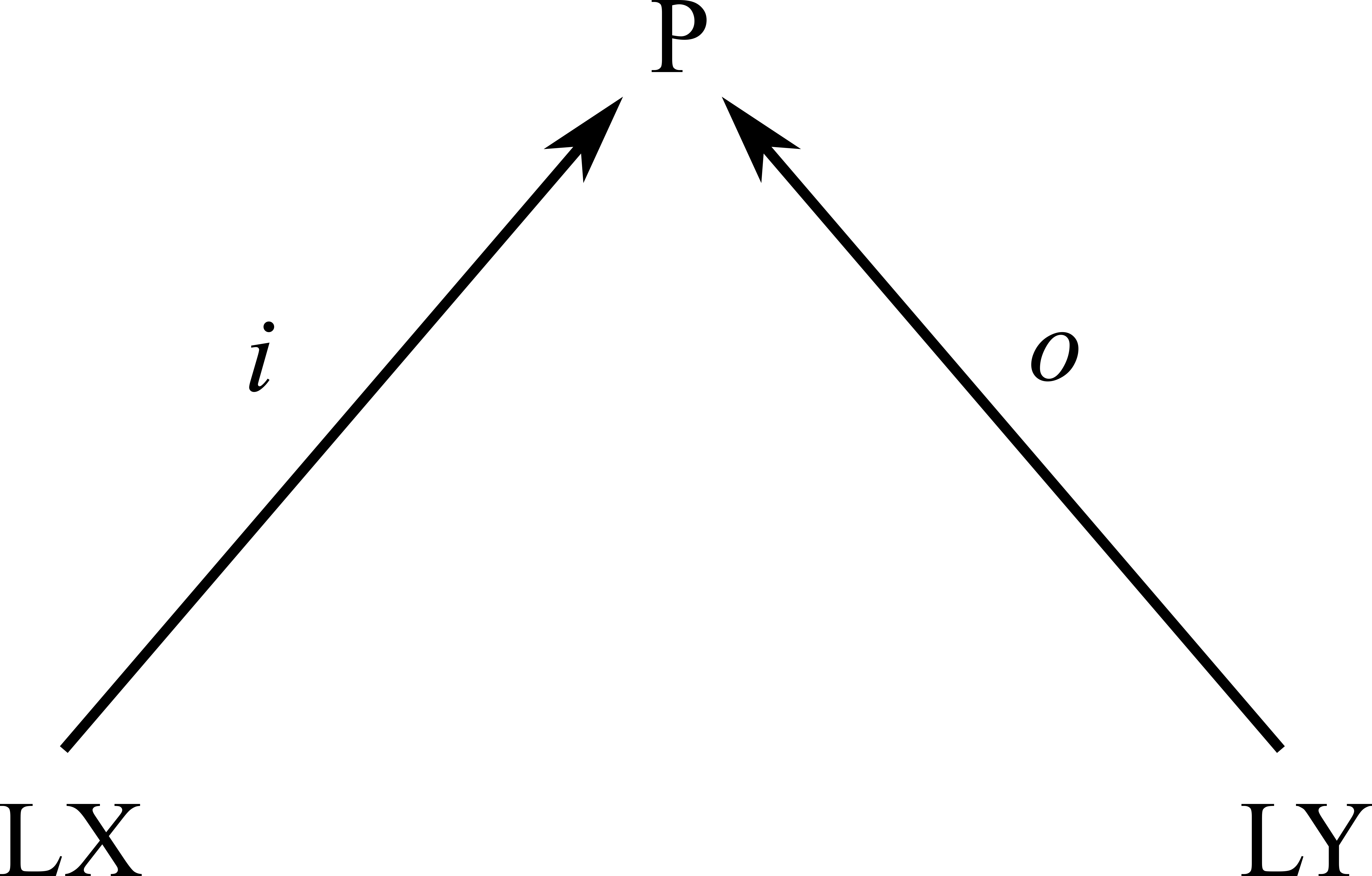}
	\label{figOpn2}
\end{figure}

\noindent
from some sets $X$ and $Y$. Here, $P$ is a Petri net, $X$ and $Y$ are the input and output set, respectively. $L: Set \rightarrow Petri$ is a left adjoint that maps a set $S$ to a Petri net $P$ with $S$ as its set of places \cite{Baez2020}. \hfill\(\Diamond\)
\end{defn}

Once ``opened'', the Petri nets (GRNs) can then be composed together. A simple modelling of GRNs with Open Petri nets can be found in \cite{Wu2019}. A more recent and much advanced modelling is established in \cite{aduddell2023compositional}.

\subsection{Using an operad to represent a gene tree} \label{operad}
Besides linear orders and networks, genes can also be organized in phylogenetic trees according to their evolutionary relationships. Interestingly, a categorical construction of phylogenetic trees has already been practised in the form of operads \cite{Baez2015}, upon which we will recapitulate some basic concepts.

Briefly, an operad is an algebraic structure where for each natural number $n = 0, 1, 2, ...$, we have a set $O_n$ whose elements are abstract n-ary operations. An element in $O_n$ can be drawn as a planar tree with one vertex connecting a root with $n$ labelled leaves.  Operations can be composed in a tree-like way to form new operations. 

When using operads to represent phylogenetic trees, we have the following definition of a phylogenetic n-tree \cite{Baez2015}:
\begin{defn}

A $\textit{phylogenetic n-tree}$ is an isomorphism class of n-trees with lengths obeying certain rules.

An n-tree for any natural number $n = 0, 1, 2, ...$ is a quadruple $T = (V, E, s, t)$ where:
\begin{itemize}
\item $V$ is a finite non-empty set whose elements are called vertices;
\item $E$ is a finite non-empty set whose elements are called edges;
\item $s: E \rightarrow V \sqcup \{1, ..., n\}$ and $t: E \rightarrow V \sqcup \{0\}$ are maps sending an edge to its source and target, respectively. For an edge $e \in E$ sending from $u$ to $v$ where $u, v \in V \sqcup \{0, 1, 2, ..., n\}$, we denote it $ u \xrightarrow{\text{e}} v $, and we have $s(e) = u, t(e) = v$;
\end{itemize}
The above data is required to satisfy the conditions:
\begin{itemize}
\item $s: E \rightarrow V \sqcup \{1, ..., n\}$ is a bijection;
\item there exists one and only one $e \in E$ such that $t(e) = 0$;
\item for any $v \in V \sqcup \{1, 2, ..., n\}$ there exist a directed path from $v$ to $0$; that is, a sequence of edges $e_i, i \in 0..n$ and vertices $v_j, j \in 1..n$ such that \\
\centerline{$v \xrightarrow{e_0} v_1$, $v_1 \xrightarrow{e_1} v_2$, ..., $v_n \xrightarrow{e_n} 0$.}
\end{itemize}
An n-tree with lengths is an n-tree together with a map $l: E \rightarrow [0, \infty)$, for any $e \in E$, we have $ l(e) $ as the length of $e$.

The rules for a phylogenetic n-tree are: \\
\hangindent=0.7cm (1) the length of every edge is positive, except for edges incident to a leaf or the root; \\
\enspace(2) there are no 0-ary or 1-ary vertices. \hfill\(\Diamond\)
\end{defn} 
\noindent
A phylogenetic tree is phylogenetic n-tree for some $n \ge 1$. Thus it provides a categorical solution for the gene trees.

\subsection{Using an olog to represent a casual activity model of gene ontology} \label{olog}
Last but not least, we introduce olog, a categorical treatment of ontology. An ontology is a formal representation of a field of knowledge. Specifically, the gene ontology (GO) provides a unified framework with controlled vocabulary to describe our biological knowledge regarding the functions of genes. Since year 2000, the Gene Ontology Consortium has been constructing three main gene ontologies accessible on its website (http://www.geneontolgy.org). They are biological process, molecular function and cellular component \cite{ashburner2000gene}.

A standard gene ontology links a gene with its individual GO terms. In order to reflect the relationships between those individual GO terms, the Gene Ontology Causal Activity Models (GO-CAM)  was developed \cite{Thomas2019}. 

Remotely related to GO, ontology log (olog) was invented as a categorical framework for knowledge representation in 2012 \cite{Spivak2012}. As its name suggests, olog was designed to record the study of ontologies. Footed in category theory, olog not only enforces mathematical rigor to ontology, but also brings abundant, comparable and scalable structures to the latter. In addition, olog has an intrinsic connection to database representation, and therefore it allows a computer-based automation of ontology storing and processing.

Following is a brief definition of an olog \cite{Spivak2012}:
\begin{defn}
An $\textit{olog}$ is a category in which each object and morphism has been labelled by text. It can be drawn as a graph of boxes with arrows connecting them.

Objects in olog are called types. A type is an abstract concept representing ``a thing" that the olog intends to describe. Each type is drawn as a box containing a singular indefinite noun phrase. 

Morphisms in olog are called aspects. An aspect of a thing, is a way of viewing or measuring that thing, and it is drawn as an arrow between two boxes. The source box is the thing that the aspect measures, and the target box gives the measuring result. Each aspect is labelled with a verb or verb phrase.

Besides types and aspects, olog includes an extra concept named fact. A fact is a way to declare that two paths with the same start and end boxes in an olog are equivalent. Facts appear as explicit equations accompany the olog they take effect in. \hfill\(\Diamond\)
 
\end{defn}

\noindent
In this definition, the purposes of types and aspects are straightforward, while that of facts are not. Intuitively, fact serves to reveal the relationships between relevant yet different measurements. Besides these simple build blocks, an olog consists of a rich repertoire of structural notions such as pullback and pushout \cite{Spivak2012}. Together, olog provides a powerful and expressive framework to organize ontologies, thus it could potentially help to improve the development of gene ontology.

We can categorify gene ontologies to ologs, and name them gene ologs. A preliminary implementations of gene ologs can be found in \cite{Wu2019b}. 

\section{An enriched category of genes}
Although we can describe different aspects of gene relationships in different sorts of categories as described in the previous section, using one category to unify them is more desirable. Therefore, we adopt an enriched category, which we name a $\textit{natural distance Lawvere metric space}$, to serve our purpose. In this section, we will first lay out the definition of that category, and then explain how it could incorporate various relations between genes. Finally, we will illustrate the enriched category with a concrete example.

\subsection{An introduction to the enriched category}
Before we explain what is an enriched category, we need to first get familiar with the concept of a symmetric monoidal preorder \cite{fong2018seven}.

\begin{defn}

A $\textit{symmetric monoid preorder}$ $(S, \le, I, \otimes)$ is a preorder $(S, \le)$ equipped with a monoidal unit $I \in S$ and a monoidal product $ \otimes: S \times S \rightarrow S $ which satisfy the following conditions:
 
\begin{itemize}
\item monotonicity: for all $x_1, x_2, y_1, y_2 \in S $, if $ x_1 \le y_1$ and $x_2 \le y_2$, then $x_1 \otimes x_2 \le y_1 \otimes y_2$;
\item unitality: for all $x \in S$, $I \otimes x = x = x \otimes I$;
\item associativity: for all $x, y, z \in S$, $(x \otimes y) \otimes z = x \otimes (y \otimes z)$;
\item symmetry: $x \otimes y = y \otimes x$.
\end{itemize}
\hfill\(\Diamond\)
\end{defn} 

\noindent
Besides, it has been proved that if $(S, \le, I, \otimes)$ is a symmetric monoidal preorder, then so is its opposite, $(S, \ge, I, \otimes)$ \cite{fong2018seven}.

Now, we are ready for the definition of an enriched category.

\begin{defn}
An $\textit{enriched category}$ $\mathcal{X}$ has as its base a symmetric monoidal preorder $\mathcal{V} = (V, \ge, I, \otimes)$, and is denoted as a $\mathcal{V}$-category. $\mathcal{X}$ consists of two components:

\begin{itemize}
\item a set $Ob(\mathcal{X})$ called objects;
\item for all $x, y \in Ob(\mathcal{X})$, we have $Hom_{\mathcal{X}}(x, y) \in V$, called the hom-object. 
\end{itemize}

And they satisfy two conditions:
\begin{itemize}
\item for all $x \in Ob(\mathcal{X})$, we have $I \ge \mathcal{X}(x, x)$;
\item for all $x, y, z \in Ob(\mathcal{X})$, we have $Hom_\mathcal{X}(x, y) \otimes Hom_\mathcal{X}(y,z) \ge Hom_\mathcal{X}(x, z)$.
\end{itemize}
\hfill\(\Diamond\)
\end{defn} 

\noindent
Intuitively, an enriched category replaces the set of morphisms (Hom-set) between two objects with an object (Hom-object) in the base category.

\subsection{An abstract category for the genes}
The symmetric monoidal preorder we use for the enriched category of genes is $ \textbf{Dist} = (\mathbb{N}, \ge, 0, +)$, where $\textbf{Dist}$ is a short name for ``distance'', $\mathbb{N}$ is the set of natural numbers, $\ge$ is the usual ordering such as $42 \ge 18$, $0$ is the monoidal unit, and $+$, the arithmetic add for natural numbers, is the monoidal product. 

Finally, we come to the definition for our category of genes:
\begin{defn}
An $\textit{enriched category}$ $\mathcal{G}$ $\textit{of genes}$ is a $\textbf{Dist}$-category which consists of:

\begin{itemize}
\item a set $Ob(\mathcal{G})$ of genes as objects;
\item for all $g_1, g_2 \in Ob(\mathcal{G})$, we have $Hom_{\mathcal{G}}(g_1, g_2) \in \mathbb{N}$, called the distance between $g_1$ and $g_2$. 
\end{itemize}

If we denote $Ob(\mathcal{G})$, the set of genes in concern for this category, as $\mathbb{G}$, and $Hom_{\mathcal{G}}(g_1, g_2)$ as $d(g_1, g_2)$, we have the the following properties:
\begin{itemize}
\item $0 \ge d(g, g) $ for all $g \in \mathbb{G}$;
\item $d(g_1, g_2) + d(g_2, g_3) \ge d(g_1, g_3) $ for all $g_1, g_2, g_3 \in \mathbb{G}$.
\end{itemize}
\hfill\(\Diamond\)
\end{defn} 

\noindent
This definition is the same as that of a Lawvere metric space \cite{fong2018seven}, except that the base is now $\textbf{Dist}$ instead of $\textbf{Cost}$. Therefore, we name it a $\textit{natural distance Lawvere metric space}$.

\subsection{Instantiations}
Now that we have settled with an abstract category $\mathcal{G}$ of genes, we can instantiate this category by putting the meaning of its morphism, the distances between genes, into the contexts of various kinds of relationships between genes.

For instance, a more concrete category of genes, say $\mathcal{G}_1$, which is a $\textbf{Dist}$-category enriched in $(\mathbb{N}, \ge, 0, +)$, could have as its objects all the genes encoded by a specific genome, denoted as $G=Ob(\mathcal{G})$. And for morphisms, if a gene $g_1$ is the closest precedent of a gene $g_2$ located on the linear form of the genome (chromosomes, see Section \ref{preorder1} for details of gene order), then there is a morphism from $g_1$ to $g_2$ with distance $1$. This category embodies the preorder relationships of genes.

Similarly, we can construct other concrete categories. They will have the same framework (all are $\textbf{Dist}$-categories) and objects as $\mathcal{G}_1$, but with different settings of morphisms. Here are a list of them:
\begin{itemize}
\item A $\textbf{Dist}$-category $\mathcal{G}_2$ representing the physical interactions between genes. For any $g_1, g_2 \in G$, if there is a direct physical interaction between them (physical interactions normally do not have directions), then we have a pair of morphisms, one goes from $g_1$ to $g_2$, and the other goes from $g_2$ to $g_1$, both with distance $1$.
\item A $\textbf{Dist}$-category $\mathcal{G}_3$ representing the genetic interactions between genes. For any $g_1, g_2 \in G$, if there is a direct genetic interaction goes from $g_1$ to $g_2$, then there is a morphism goes from $g_1$ to $g_2$, with distance $1$. Note that genetic interaction does have directions. More importantly, there are two types of genetic interactions, one is suppression and the other is enhancement. In $\mathcal{G}_3$, we keep the direction but ignore the type of a morphism. The benefit for this choice is to keep the morphisms composable, and the price is a loss of the type information.
\item A $\textbf{Dist}$-category $\mathcal{G}_4$ representing the signalling pathways of genes. For any $g_1, g_2 \in G$, if $g_1$ is the direct upstream component of $g_2$ in a specific signalling pathway, then there is a morphism goes from $g_1$ to $g_2$. Note that different signalling pathways have many component genes in common, but we did not foresee any serious distortions of known biological facts by blurring the distinct signalling pathways in the same category, at least at this moment.
\item A $\textbf{Dist}$-category $\mathcal{G}_5$ representing the gene groups. For any $g_1, g_2 \in G$, if $g_1$ is in the same gene group as $g_2$, then we have a pair of morphisms, one goes from $g_1$ to $g_2$, and the other goes from $g_2$ to $g_1$, both with distance $1$.
\item A $\textbf{Dist}$-category $\mathcal{G}_6$ representing the gene interaction in the 3D-genome structure. For any $g_1, g_2 \in G$, if there is a direct interaction between $g_1$ and $g_2$ in the 3D-genome structure, then we have a pair of morphisms, one goes from $g_1$ to $g_2$, and the other goes from $g_2$ to $g_1$, both with distance $1$.
\end{itemize}

For any category $\mathcal{G}_i$, $i \in 1..6 $, we define the following:
\begin{itemize}
\item distance for identity morphism: $ d(g, g) = 0 $ for all $g \in \mathbb{G}_i$;
\item distance for morphism composition: $ d(g_1, g_3) = d(g_1, g_2) + d(g_2, g_3) $ for all $g_1, g_2, g_3 \in \mathbb{G}_i$.
\end{itemize}

\noindent
Intuitively, for any two genes in a specific $\textbf{Dist}$-category, if there is no relationship between the two genes, then there is no path connecting their corresponding nodes in the graph of the category. Otherwise, there is at least one path connecting the two genes. 

\subsection{A concrete example}

\noindent 
To provide a concrete example of these categories, we will be using the fly genome release 6.50, which contains a total of 17,896 genes \cite{hoskins2015release}. Although our ideal approach would involve analyzing the entire gene set, we have chosen to focus on a specific subset of genes that belong to the classical MAPK signaling pathway. This decision was made to facilitate visualization and simplify our analysis.

First, we illustrate the MAPK pathway genes in their cellular context in Figure \ref{figMAPKdiagram}:

\begin{figure}[H]
	\centering
	\includegraphics[scale=0.4]{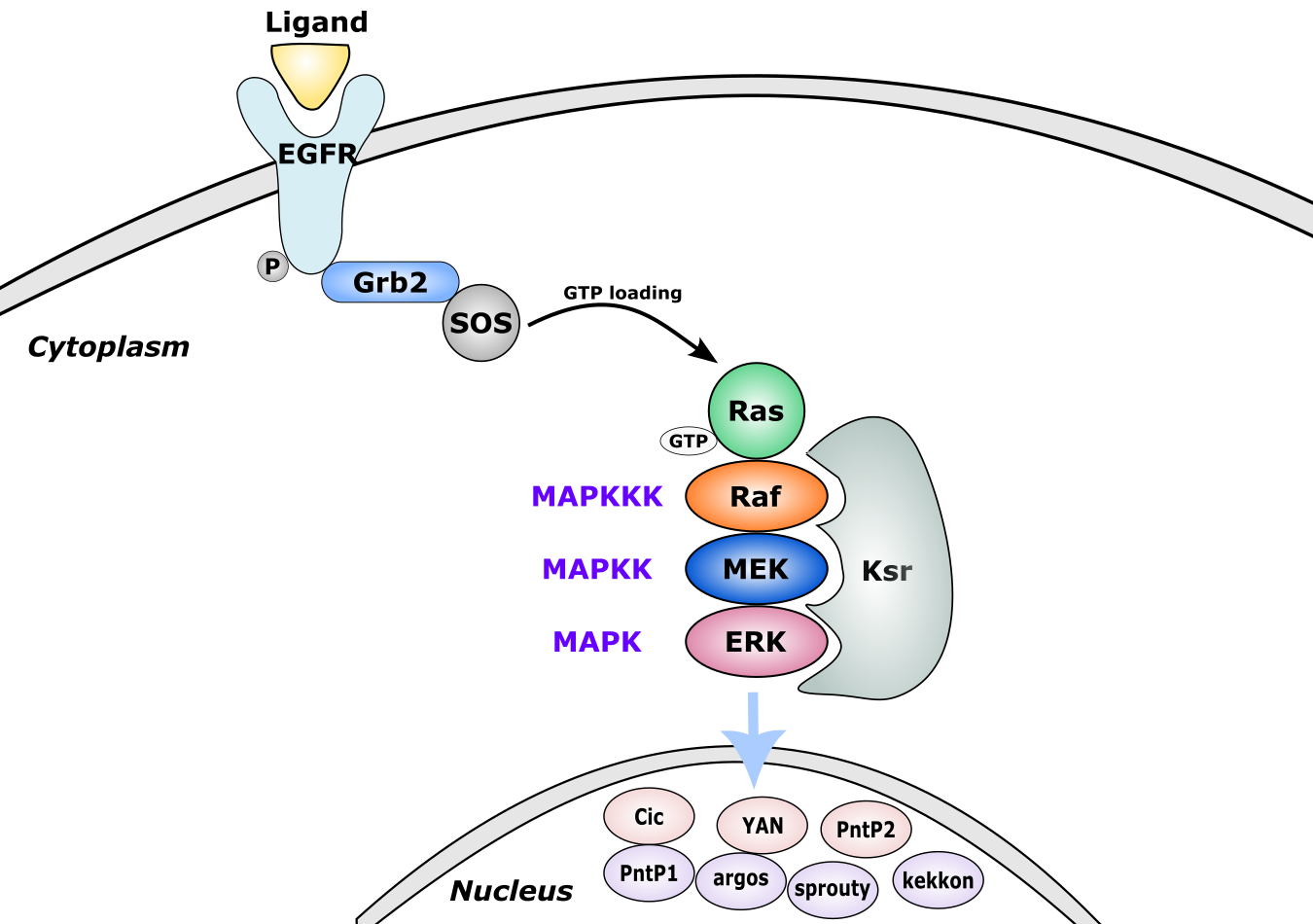}
	\captionsetup{justification=centering}
	\caption{The MAPK pathway of \textit{Drosophila melanogaster} \\ (adapted from \cite{shilo2014regulation})}
	\label{figMAPKdiagram}
\end{figure}

\noindent
The relevant genes are listed in Table \ref{tabMAPKgenes}:

\begin{table}[h!]
\centering
\scalebox{0.8}{
\begin{tabular}{|p{2cm}|p{2cm}|p{1cm}|p{2cm}|p{2cm}|} 
 \hline
 \textbf{Component} & \textbf{Gene name} & \textbf{Chr} & \textbf{Start} & \textbf{Stop} \\
 \hline
 EGFR & Egfr & 2R & 21,522,420 & 21,559,977 \\
 \hline
 Grb2 & drk & 2R & 13,495,222 & 13,502,841 \\
 \hline
 SOS & Sos & 2L & 13,813,816 & 13,819,824 \\
 \hline
 RAS & Ras85D & 3R & 9,510,561 & 9,513,067 \\
 \hline
 Ksr & ksr & 3R & 5,478,390 & 5,483,906 \\
 \hline
 Raf & Raf & X & 2,295,466 & 2,343,870 \\
 \hline
 MEK & Dsor1 & X & 9,247,342 & 9,250,037 \\
 \hline
 ERK & rl & 2R & 1,071,462 & 1,125,927 \\
 \hline
 YAN & aop & 2L & 2,156,484 & 2,178,754 \\
 \hline
 PntP1 & pnt & 3R & 23,290,231 & 23,346,167 \\
 \hline
 PntP2 & pnt & 3R & 23,290,231 & 23,346,167 \\
 \hline 
 Cic & cic & 3R & 20,252,770 & 20,303,942 \\
 \hline 
 argos & aos & 3L & 16,470,386 & 16,483,650 \\
 \hline 
 sprouty & sty & 3L & 3,401,153 & 3,424,935 \\
 \hline 
 kekkon & kek1 & 2L & 12,817,000 & 12,822,787 \\ 
 \hline
\end{tabular}}
\caption{The MAPK pathway genes}
\label{tabMAPKgenes}
\end{table}

\noindent
Next, we show the positions of the MAPK pathway genes on the ideogram of the fly genome in Figure \ref{figMAPKlig}:

\begin{figure}[H]
	\centering
	\includegraphics[scale=0.3]{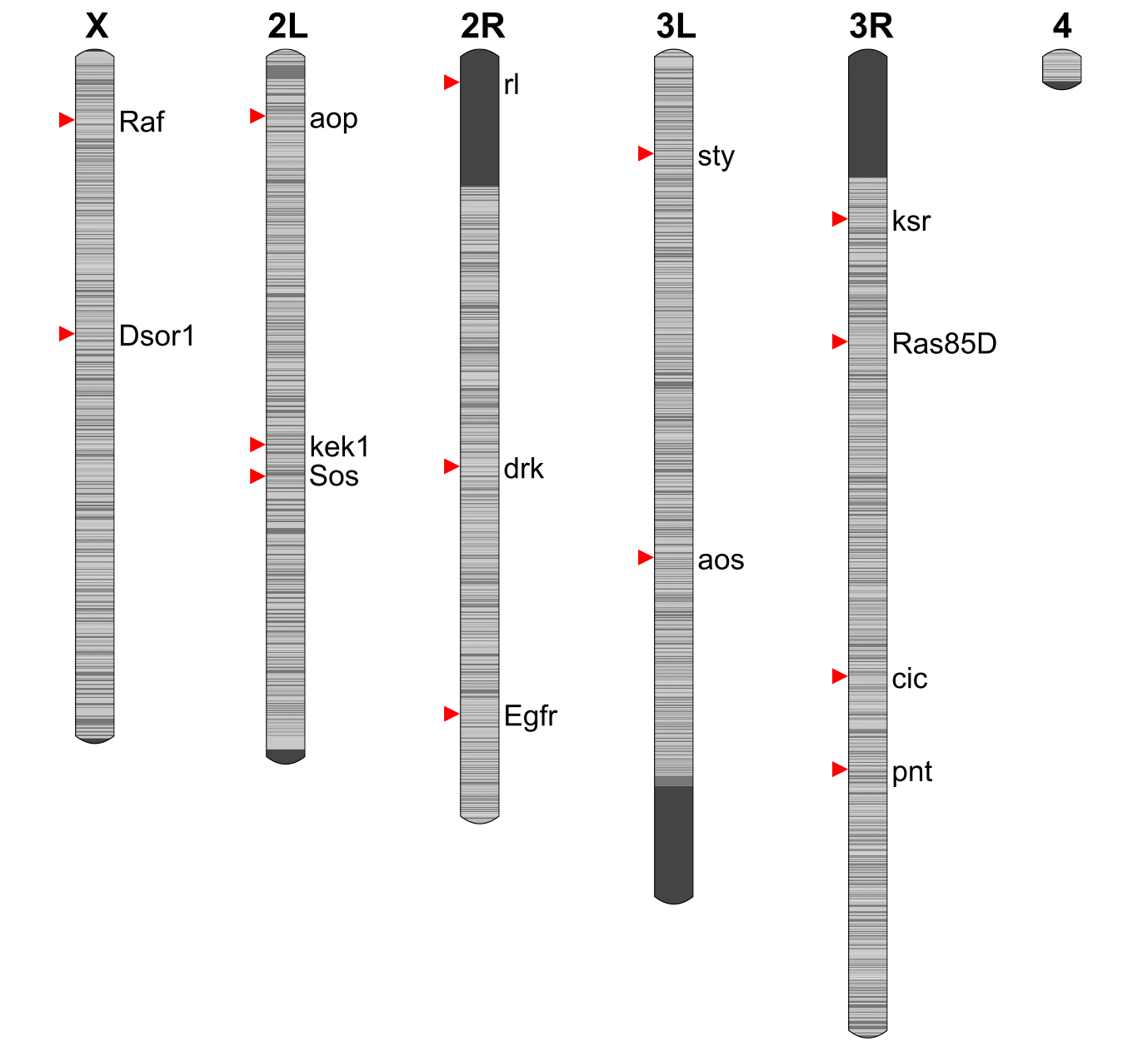}
	\captionsetup{justification=centering}
	\caption{The MAPK pathway genes on \textit{Drosophila} ideogram (linear) \\ (created via https://eweitz.github.io/ideogram/)}
	\label{figMAPKlig}
\end{figure}

\begin{figure}[H]
	\centering
	\includegraphics[scale=0.4]{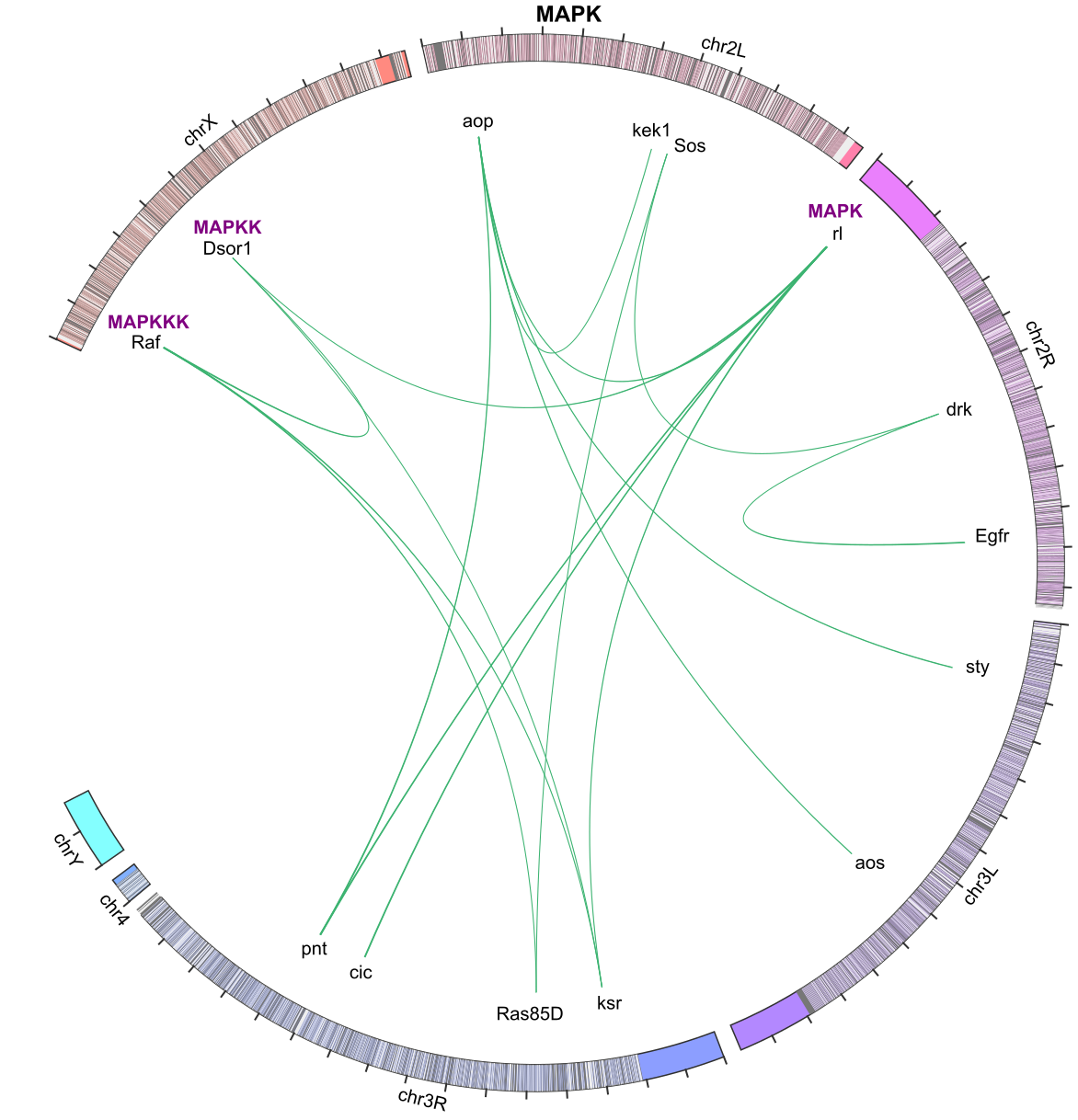}
	\captionsetup{justification=centering}
	\caption{The MAPK pathway genes on \textit{Drosophila} ideogram (circular) \\ (created via https://github.com/ponnhide/pyCircos)}
	\label{figMAPKcig}
\end{figure}

\noindent
Due to the complexity of visualizing gene interactions on a linear ideogram, we have opted to present the MAPK genes on a circular ideogram in Figure \ref{figMAPKcig}. To illustrate the signaling pathway depicted in Figure \ref{figMAPKdiagram}, we have included a curve that links the genes known to interact with each other.

\noindent
Now, we distil two categories from the above information, Figure \ref{figMAPKcatGO} shows the first one, with its morphisms encode the gene orders on the chromosomes:


\begin{figure}[H]
\begin{center}
\begin{tikzcd}[boxed]
\begin{array}{c} Raf \\ \bullet \end{array} \arrow[r, "1", bend left] & \begin{array}{c} Dsor1 \\ \bullet \end{array}                            &                                                                        &                                              \\
\begin{array}{c} aop \\ \bullet \end{array} \arrow[r, "1", bend left] & \begin{array}{c} kek1 \\ \bullet \end{array} \arrow[r, "1", bend left]   & \begin{array}{c} Sos \\ \bullet \end{array}                           &                                              \\
\begin{array}{c} rl \\ \bullet \end{array} \arrow[r, "1", bend left]  & \begin{array}{c} drk \\ \bullet \end{array} \arrow[r, "1", bend left]    & \begin{array}{c} Egfr \\ \bullet \end{array}                          &                                              \\
\begin{array}{c} sty \\ \bullet \end{array} \arrow[r, "1", bend left] & \begin{array}{c} aos \\ \bullet \end{array}                              &                                                                        &                                              \\
\begin{array}{c} ksr \\ \bullet \end{array} \arrow[r, "1", bend left] & \begin{array}{c} Ras85D \\ \bullet \end{array} \arrow[r, "1", bend left] & \begin{array}{c} cic \\ \bullet \end{array} \arrow[r, "1", bend left] & \begin{array}{c} pnt \\ \bullet \end{array}
\end{tikzcd}
\end{center}
\captionsetup{justification=centering}
\caption{A category of MAPK genes representing the gene orders \\ (all commutative diagrams are created via https://tikzcd.yichuanshen.de/)}
\label{figMAPKcatGO}
\end{figure}
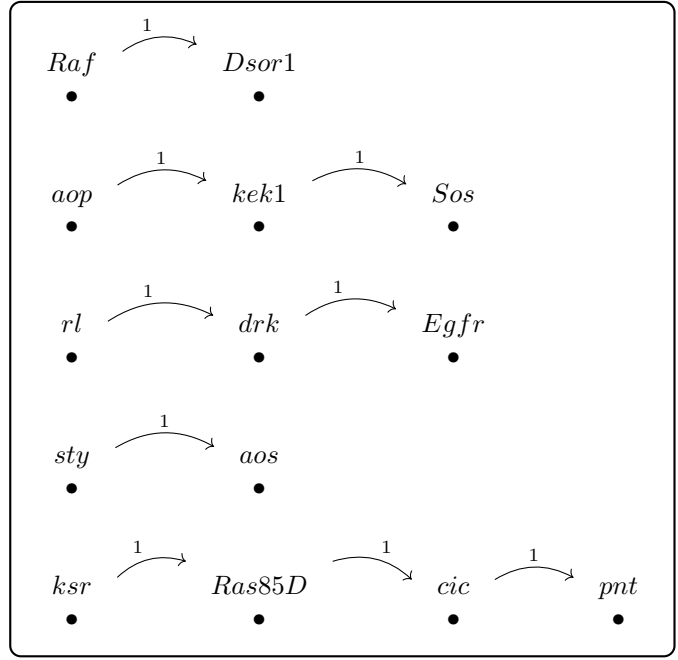

\noindent
In this category diagram, we have omitted identity morphisms to avoid clutter. In Figure \ref{figMAPKsub} we illustrate the identity morphisms as well as a composition morphism for a subset of the MAPK genes:

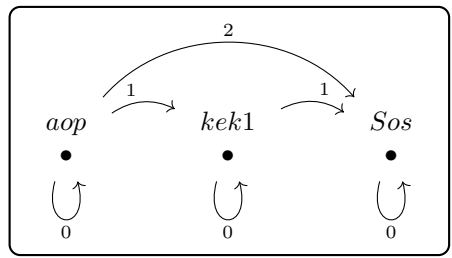
\begin{figure}[H]
\begin{center}
\begin{tikzcd}[boxed]
\begin{array}{c} aop \\ \bullet  \arrow["0"', loop, distance=2em, in=285, out=255] \end{array} \arrow[r, "1", bend left] \arrow[rr, "2", bend left=49] & \begin{array}{c} kek1 \\ \bullet  \arrow["0"', loop, distance=2em, in=285, out=255] \end{array} \arrow[r, "1", bend left] & \begin{array}{c} Sos \\ \bullet \arrow["0"', loop, distance=2em, in=285, out=255] \end{array}
\end{tikzcd}
\end{center}
\caption{A category of 3 MAPK genes on chromosome 2L}
\label{figMAPKsub}
\end{figure}

\noindent
The second category shown in Figure \ref{figMAPKcatSP} embodies the gene interactions in the MAPK signalling pathway:

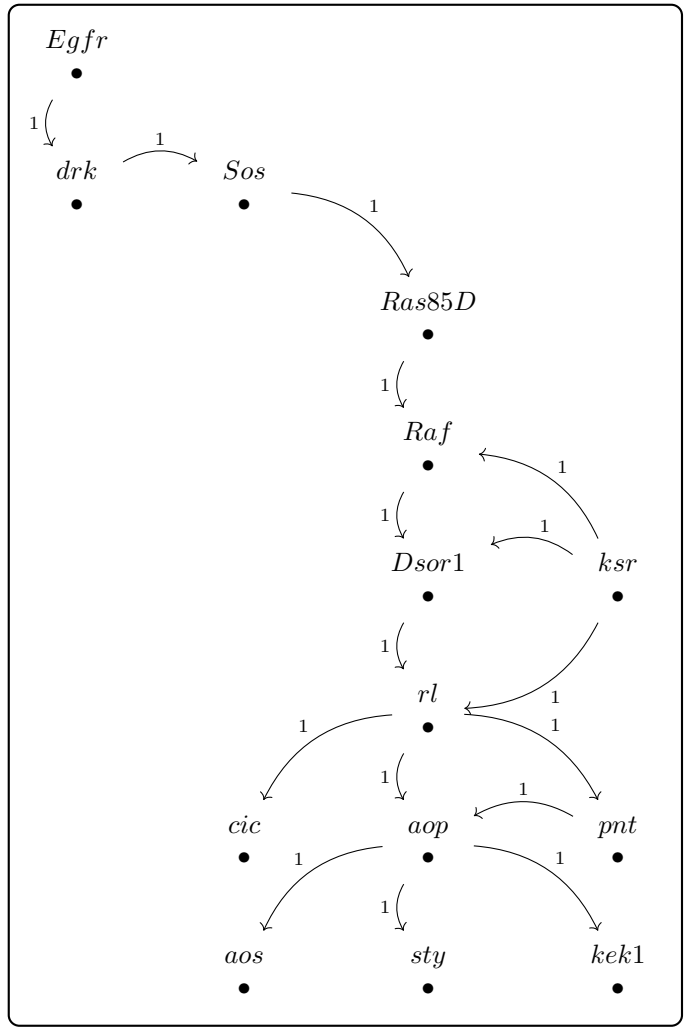
\begin{figure}[H]
\begin{center}
\begin{tikzcd}[boxed]
\begin{array}{c} Egfr \\ \bullet \end{array} \arrow[d, "1"', bend right] &                                                                        &                                                                                                                                 &                                                                                                                                 \\
\begin{array}{c} drk \\ \bullet \end{array} \arrow[r, "1", bend left]    & \begin{array}{c} Sos \\ \bullet \end{array} \arrow[rd, "1", bend left] &                                                                                                                                 &                                                                                                                                 \\
                                                                         &                                                                        & \begin{array}{c} Ras85D \\ \bullet \end{array} \arrow[d, "1"', bend right]                                                      &                                                                                                                                 \\
                                                                         &                                                                        & \begin{array}{c} Raf \\ \bullet \end{array} \arrow[d, "1"', bend right]                                                         &                                                                                                                                 \\
                                                                         &                                                                        & \begin{array}{c} Dsor1 \\ \bullet \end{array} \arrow[d, "1"', bend right]                                                       & \begin{array}{c} ksr \\ \bullet \end{array} \arrow[lu, "1"', bend right] \arrow[l, "1"', bend right] \arrow[ld, "1", bend left] \\
                                                                         &                                                                        & \begin{array}{c} rl \\ \bullet \end{array} \arrow[ld, "1"', bend right] \arrow[d, "1"', bend right] \arrow[rd, "1", bend left]  &                                                                                                                                 \\
                                                                         & \begin{array}{c} cic \\ \bullet \end{array}                            & \begin{array}{c} aop \\ \bullet \end{array} \arrow[ld, "1"', bend right] \arrow[d, "1"', bend right] \arrow[rd, "1", bend left] & \begin{array}{c} pnt \\ \bullet \end{array} \arrow[l, "1"', bend right]                                                         \\
                                                                         & \begin{array}{c} aos \\ \bullet \end{array}                            & \begin{array}{c} sty \\ \bullet \end{array}                                                                                     & \begin{array}{c} kek1 \\ \bullet \end{array}                                                                                   
\end{tikzcd}
\end{center}
\caption{A category of MAPK genes representing the MAPK pathway}
\label{figMAPKcatSP}
\end{figure}

\noindent
With this example, we showcase the capability of $\textbf{Dist}$-category to describe different gene relations in a consistent framework.

\section{Conclusion and future work}

In this manuscript, we briefly review several existing gene categories and introduce a novel category called $\textbf{Dist}$-category to model gene relationships. 

Our work offers two key contributions: (1) $\textbf{Dist}$-category provides a unified framework for describing various types of gene relationships. (2) By utilizing distance as an abstraction to represent gene relationships, $\textbf{Dist}$-category achieves a biologically meaningful composition of morphisms.

Through this foundational gene category, we can explore various constructions from category theory, such as product (coproduct) or pullback (pushout). By doing so, we will potentially uncover novel biological insights regarding how the genome encodes gene relationships.

\bibliographystyle{apalike}
\bibliography{CatGenes}

\begin{thebibliography}{}

\bibitem[Aduddell et~al., 2023]{aduddell2023compositional}
Aduddell, R., Fairbanks, J., Kumar, A., Ocal, P.~S., Patterson, E., and
  Shapiro, B.~T. (2023).
\newblock A compositional account of motifs, mechanisms, and dynamics in
  biochemical regulatory networks.
\newblock {\em arXiv preprint arXiv:2301.01445}.

\bibitem[Ashburner et~al., 2000]{ashburner2000gene}
Ashburner, M., Ball, C.~A., Blake, J.~A., Botstein, D., Butler, H., Cherry,
  J.~M., Davis, A.~P., Dolinski, K., Dwight, S.~S., Eppig, J.~T., et~al.
  (2000).
\newblock Gene ontology: tool for the unification of biology.
\newblock {\em Nature genetics}, 25:25.

\bibitem[Awodey, 2010]{awodey2010category}
Awodey, S. (2010).
\newblock {\em Category theory}.
\newblock Oxford University Press.

\bibitem[Baez and Master, 2020]{Baez2020}
Baez, J.~C. and Master, J. (2020).
\newblock Open petri nets.
\newblock {\em Mathematical Structures in Computer Science}, 30:314--341.

\bibitem[Baez and Otter, 2015]{Baez2015}
Baez, J.~C. and Otter, N. (2015).
\newblock Operads and phylogenetic trees.
\newblock {\em Theory and Applications of Categories}, 32:1397--1453.

\bibitem[Bordon and Mraz, 2012]{Bordon2012}
Bordon, J. and Mraz, M. (2012).
\newblock Modeling gene regulatory networks using petri nets.

\bibitem[Coecke and Kissinger, 2017]{Coecke2017}
Coecke, B. and Kissinger, A. (2017).
\newblock Picturing quantum processes: A first course in quantum theory and
  diagrammatic reasoning.
\newblock {\em Picturing Quantum Processes: A First Course in Quantum Theory
  and Diagrammatic Reasoning}, pages 1--844.

\bibitem[Fong and Spivak, 2018]{fong2018seven}
Fong, B. and Spivak, D.~I. (2018).
\newblock Seven sketches in compositionality: An invitation to applied category
  theory.
\newblock {\em arXiv preprint arXiv:1803.05316}.

\bibitem[Hoskins et~al., 2015]{hoskins2015release}
Hoskins, R.~A., Carlson, J.~W., Wan, K.~H., Park, S., Mendez, I., Galle, S.~E.,
  Booth, B.~W., Pfeiffer, B.~D., George, R.~A., Svirskas, R., et~al. (2015).
\newblock The release 6 reference sequence of the drosophila melanogaster
  genome.
\newblock {\em Genome research}, 25(3):445--458.

\bibitem[Karlebach and Shamir, 2008]{Karlebach2008}
Karlebach, G. and Shamir, R. (2008).
\newblock Modelling and analysis of gene regulatory networks.
\newblock {\em Nature Reviews Molecular Cell Biology 2008 9:10}, 9:770--780.

\bibitem[Milewski and Tabachnik, 2019]{Milewski2019}
Milewski, B. and Tabachnik, I. (2019).
\newblock {\em Category theory for programmers}.

\bibitem[Murata, 1989]{Murata1989}
Murata, T. (1989).
\newblock Petri nets: Properties, analysis and applications.
\newblock {\em Proceedings of the IEEE}, 77:541--580.

\bibitem[Petri, 1962]{petri1962kommunikation}
Petri, C.~A. (1962).
\newblock Kommunikation mit automaten.

\bibitem[Reisig, 1985]{Reisig1985}
Reisig, W. (1985).
\newblock {\em Petri Nets : an Introduction}.
\newblock Springer Berlin Heidelberg.

\bibitem[Riehl, 2017]{riehl2017category}
Riehl, E. (2017).
\newblock {\em Category theory in context}.
\newblock Courier Dover Publications.

\bibitem[Shilo, 2014]{shilo2014regulation}
Shilo, B.-Z. (2014).
\newblock The regulation and functions of mapk pathways in drosophila.
\newblock {\em Methods}, 68(1):151--159.

\bibitem[Sica, 2006]{sica2006category}
Sica, G. (2006).
\newblock {\em What is category theory?}, volume~3.
\newblock Polimetrica sas.

\bibitem[Southwell and Gupta, 2021]{Southwell2021}
Southwell, R. and Gupta, N. (2021).
\newblock {\em Categories and Toposes: Visualized and Explained}.

\bibitem[Spivak, 2014]{spivak2014category}
Spivak, D.~I. (2014).
\newblock {\em Category theory for the sciences}.
\newblock MIT Press.

\bibitem[Spivak and Kent, 2012]{Spivak2012}
Spivak, D.~I. and Kent, R.~E. (2012).
\newblock Ologs: A categorical framework for knowledge representation.
\newblock {\em PLoS ONE}, 7:1--52.

\bibitem[Steggles et~al., 2007]{Steggles2007}
Steggles, L.~J., Banks, R., Shaw, O., and Wipat, A. (2007).
\newblock Qualitatively modelling and analysing genetic regulatory networks: a
  petri net approach.
\newblock {\em Bioinformatics (Oxford, England)}, 23:336--343.

\bibitem[Thomas et~al., 2019]{Thomas2019}
Thomas, P.~D., Hill, D.~P., Mi, H., Osumi-Sutherland, D., Auken, K.~V., Carbon,
  S., Balhoff, J.~P., Albou, L.~P., Good, B., Gaudet, P., Lewis, S.~E., and
  Mungall, C.~J. (2019).
\newblock Gene ontology causal activity modeling (go-cam) moves beyond go
  annotations to structured descriptions of biological functions and systems.
\newblock {\em Nature Genetics 2019 51:10}, 51:1429--1433.

\bibitem[Tuyéras, 2018a]{Tuyeras2018}
Tuyéras, R. (2018a).
\newblock Category theory for genetics i: mutations and sequence alignments.

\bibitem[Tuyéras, 2018b]{Tuyeras2018a}
Tuyéras, R. (2018b).
\newblock Category theory for genetics ii: genotype, phenotype and haplotype.

\bibitem[Wu, 2019a]{Wu2019b}
Wu, Y. (2019a).
\newblock Gene ologs: a categorical framework for gene ontology,
  arxiv:1909.11210 [q-bio.gn].

\bibitem[Wu, 2019b]{Wu2019}
Wu, Y. (2019b).
\newblock An open petri net implementation of gene regulatory networks,
  arxiv:1907.11316 [q-bio.mn].

\bibitem[Wu, 2020]{Wu2020}
Wu, Y. (2020).
\newblock Applied category theory for genomics-an initiative.

\end{thebibliography}

\end{document}